\documentclass[twocolumn,aps,showpacs]{revtex4}
\usepackage{amsmath}
\usepackage{amssymb}
\usepackage{graphicx}
\usepackage{dcolumn}

\begin{document}

\title{Interference oscillations of microwave photoresistance
in double quantum wells}
\author{S. Wiedmann$^{1,3}$, G. M. Gusev,$^2$ O. E. Raichev,$^{2}$\footnote{Permanent address: 
Institute of Semiconductor Physics, Prospekt Nauki 45, 03028, Kiev, Ukraine} T. E. Lamas,$^2$ A. K.
Bakarov,$^{2}$\footnote{Permanent address: Institute of Semiconductor Physics, Novosibirsk 630090, Russia} 
and J. C. Portal$^{1,3,4}$}
\affiliation{$^1$GHMFL-CNRS, BP-166, F-38042, Grenoble, Cedex 9,
France} \affiliation{$^2$Instituto de F\'{\i}sica da Universidade de
S\~ao Paulo, CP 66318 CEP 05315-970, S\~ao Paulo, SP, Brazil}
\affiliation{$^3$INSA-Toulouse, 31077, Cedex 4, France}
\affiliation{$^4$Institut Universitaire de France, Toulouse, France}

\begin{abstract}
We observe oscillatory magnetoresistance
in double quantum wells under microwave irradiation. The results are
explained in terms of the influence of subband coupling on the
frequency-dependent photoinduced part of the electron distribution
function. As a consequence, the magnetoresistance demonstrates the
interference of magneto-intersubband oscillations and conventional
microwave-induced resistance oscillations.


\end{abstract}

\normalsize\date{\today}

\maketitle

The phenomenon of microwave-induced resistance oscillations$^{1-3}$
(MIRO) in two-dimensional (2D) electron systems under perpendicular
magnetic fields has attracted much interest.$^{4}$ These
oscillations are periodic in the inverse magnetic field with a
period determined by the ratio of the microwave radiation frequency
$\omega$ to the cyclotron frequency $\omega_c$ and survive at high
temperatures. Basically, the observed oscillatory photoconductivity
is caused by the Landau quantization of electron states, though
different microscopic mechanisms of this phenomenon are still under
discussion.

The influence of microwave irradiation on the magnetotransport
properties is currently under investigation in quantum wells with a
single occupied subband. In our paper, we underline the importance
of similar studies for the systems with two occupied 2D subbands,
where the magnetotransport shows special new features as
compared to the single-subband case. Apart from the commonly known
Shubnikov-de Haas oscillations (SdHO), there exist the
magneto-intersubband (MIS) oscillations of resistivity caused by
periodic modulation of the probability of intersubband transitions
by the magnetic field$^{5,6}$ (see Ref. 7 for more references).
These oscillations survive at high temperatures, because they are
not related to the position of the Landau levels with respect to the
Fermi surface. 
\\Our recent observation$^{7}$ of large-amplitude MIS
oscillations at magnetic fields below 1 T in high-mobility double
quantum wells (DQWs) has established that these oscillations are a
well-reproducible feature of magnetotransport in such systems. The
present measurements, supported by a theoretical analysis, suggest
that the behavior of oscillating magnetoresistance in DQWs under
microwave irradiation is caused by an interference of the physical
mechanisms responsible for the MIS oscillations and conventional
MIRO. We have studied dependence of the resistance of symmetric balanced
GaAs DQWs with wells widths of 14 nm and different barrier widths $d_{b}=1.4$, 
2, and 3 nm on the magnetic field $B$ in the presence of microwave irradiation 
of different frequencies and at different temperatures and intensities of radiation.

\begin{figure}[ht]
\begin{center}\leavevmode
\includegraphics[width=9cm]{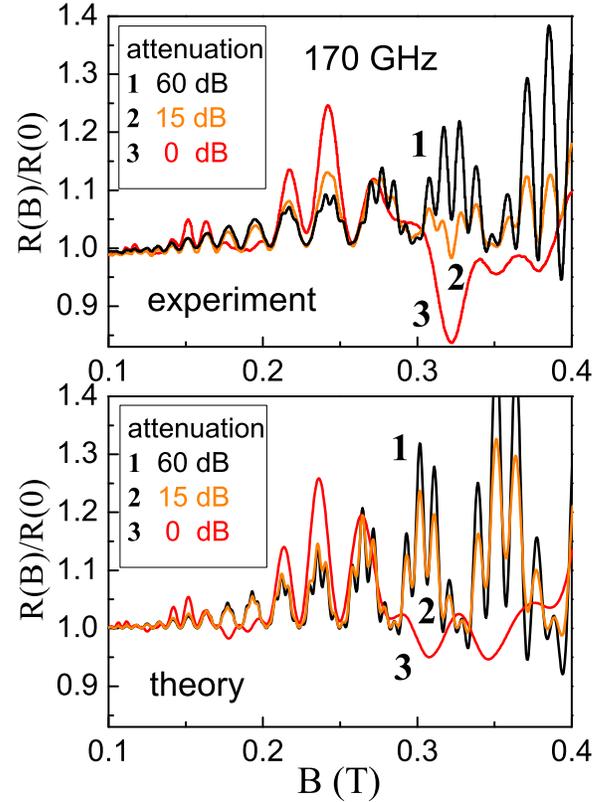}
\end{center}
\addvspace{-0.8 cm} \caption{(Color online) Measured (upper panel)
and calculated (lower panel) magnetoresistance of DQWs for different
intensities of 170 GHz radiation at $T=1.4$ K. The effect of the
radiation can be neglected for the attenuation of 60 dB.}
\end{figure}

 The samples have high
mobility of $10^{6}$ cm$^{2}$/V s and high total sheet electron
density $n_s \simeq 10^{12}$ cm$^{-2}$. The samples were mounted in
a waveguide with different cross sections. The measurements were
performed for perpendicular and parallel orientation of the current
with respect to microwave polarization. The resistance $R=R_{xx}$
was measured by using the standard low-frequency lock-in technique.
No polarization dependence of the resistance has been observed.
While similar results were obtained for samples with different
barrier widths, the data reported here corresponds to 1.4 nm barrier
samples. Without the irradiation, the magnetoresistance shows
large-period MIS oscillations clearly visible starting from $B=0.1$
T, and it also shows small period SdHO superimposed on the MIS
oscillation pattern at higher fields$^{7}$ (see Fig. 1). The subband separation in our
structure, $\Delta_{12}=3.67$ meV, has been found from the MIS
oscillation periodicity at low $B$. 

\begin{figure}[ht]
\begin{center}\leavevmode
\includegraphics[width=9cm]{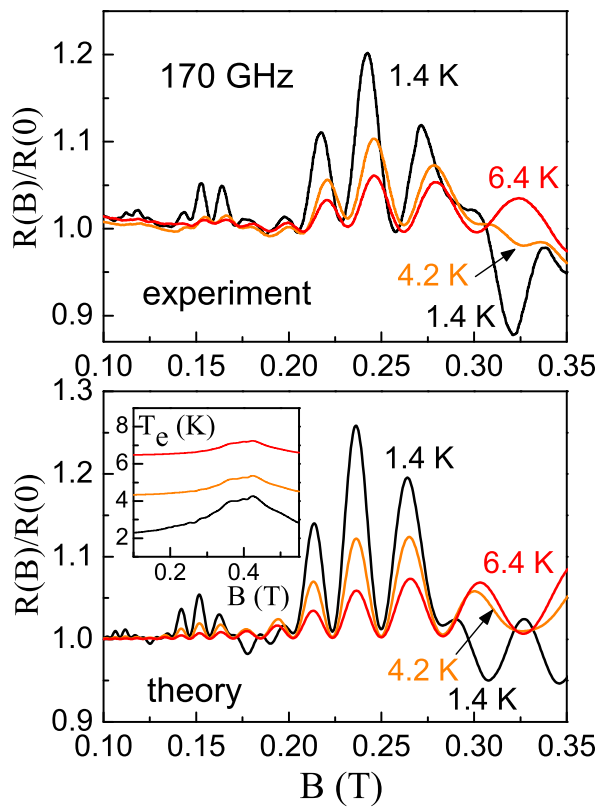}
\end{center}
\addvspace{-0.8 cm} \caption{(Color online) Measured (upper panel)
and calculated (lower panel) magnetoresistance of DQWs under 170 GHz
irradiation for different temperatures. The inset shows the
dependence of electron temperature $T_e$ on the magnetic field. Weak
oscillations of $T_e$ are caused by oscillations of the absorbed
radiation power.}
\end{figure}

An increase in the intensity of
the radiation leads to the expected damping of the SdHO owing to
electron heating, and also causes suppression of some groups of MIS
peaks, followed by the inversion (flip) of these peaks at high
intensity, while the other MIS peaks are continuously enhanced by
the irradiation. With increasing temperature, the inverted peaks go
up and the enhanced peaks go down, so the conventional MIS
oscillation picture is restored (Fig. 2). Measurements at different
frequencies (Fig. 3) have confirmed that the selective peak flip is
correlated with the radiation frequency and follows the periodicity
determined by the ratio $\omega/\omega_c$. This allows us to
attribute the observed effect to MIRO-related phenomena. However,
the peak flip cannot be explained by a simple superposition of the
factors $\cos (2 \pi \Delta_{12}/\hbar \omega_c)$ and $-\sin (2 \pi
\omega/ \omega_c)$ describing the MIS oscillations and the MIRO,
respectively.

\begin{figure}[ht]
\begin{center}\leavevmode
\includegraphics[width=9.5cm]{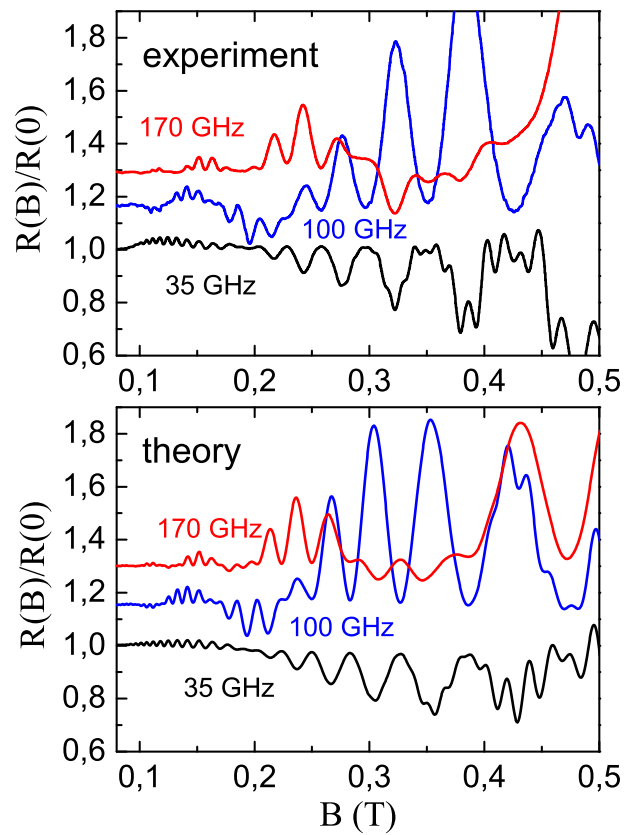}
\end{center}
\addvspace{-0.8 cm} \caption{(Color online) Measured (upper panel)
and calculated (lower panel) magnetoresistance of DQWs at $T=1.4$ K
for different frequencies of microwave excitation. The curves for
100 and 170 GHz are shifted up for clarity.}
\end{figure}

The theoretical explanation of our data is based on the physical
model of Dmitriev {\em et al.},$^{8}$ generalized to the two-subband
case and improved by taking into account electrodynamic effects in
the absorption of radiation by 2D layers.$^{9-11}$ We assume that
the elastic scattering of electrons, including the intersubband
scattering, is much stronger than the inelastic one, so the electron
distribution function (averaged over the time period $2 \pi/\omega$)
under microwave excitation remains quasi-isotropic and common for
both subbands, though essentially non-equilibrium. This energy
distribution function, $f_{\varepsilon}$, is found from the kinetic
equation $G_{\varepsilon}(f)=-J_{\varepsilon}(f)$, where
$G_{\varepsilon}$ is the generation term, and $J_{\varepsilon}$ is
the inelastic collision integral. Under the excitation by linearly
polarized high-frequency electric field ${\bf E}_t={\bf E}
\cos(\omega t)$, weak enough to neglect the multi-photon processes,
the generation term is written as
\begin{eqnarray}
G_{\varepsilon}=\frac{1}{D_{\varepsilon}} \frac{e^2}{8 \pi m
\omega^2} \sum_{\pm} |E^{(\pm)}_{\omega}|^2  \left[
\Phi^{(\pm)}_{\varepsilon}(\omega) (f_{\varepsilon+\hbar
\omega}-f_{\varepsilon}) \right. \nonumber \\
\left.+ \Phi^{(\pm)}_{\varepsilon-\hbar \omega}(\omega)
(f_{\varepsilon-\hbar \omega}-f_{\varepsilon}) \right].
%1
\end{eqnarray}
where $D_{\varepsilon}$ is the density of states, $e$ is the
electron charge, and $m$ is the effective mass of electron. Next,
$E^{(\pm)}_{\omega} = E \lambda^{(\pm)}_{\omega}$, where
$\lambda^{(\pm)}_{\omega}=[1+2 \pi \sigma_{\pm}(\omega)/c
\sqrt{\epsilon}]^{-1}$, are the amplitudes of the circularly
polarized components of electric field in the 2D plane, found from
the Maxwell equations.$^{9,10}$ Here, $\sigma_{\pm}(\omega)$ are the
complex conductivities describing response of the electron system to
the circularly polarized fields, $c$ is the velocity of light, and
$\epsilon$ is the dielectric permittivity of the medium surrounding
the quantum well. The functions $\Phi^{(\pm)}_{\varepsilon}(
\omega)$, which describe the probability of electron transitions
between the states with energies $\varepsilon$ and $\varepsilon
+\hbar \omega$, are written in terms of Green's functions as
$\Phi^{(\pm)}_{\varepsilon}(\omega)={\rm Re}
(Q^{AR(\pm)}_{\varepsilon,\omega}-Q^{AA(\pm)}_{\varepsilon,\omega})$,
where
\begin{eqnarray}
Q_{\varepsilon,\omega}^{ss'(+)} = \frac{2 \omega_c}{L^2}
\sum_{nn'} \sum_{jj'} \sqrt{(n+1)(n'+1)} \sum_{p_y p_y'}  \nonumber \\
\times \left< \left< G^{jj',s}_{\varepsilon}(n+1 p_y, n'+1 p'_y)
G^{j'j,s'}_{\varepsilon+\hbar \omega}(n' p'_y, n p_y) \right>
\right>,
%2
\end{eqnarray}
and $Q_{\varepsilon,\omega}^{ss'(-)}$ is given by the permutation of
the indices, $n+1 \leftrightarrow n$ and $n'+1 \leftrightarrow n'$,
in the arguments of Green's functions in Eq. (2). The Green's
functions $G$, retarded (R) and advanced (A), are determined by
interaction of electrons with static disorder potential in the
presence of magnetic field. They are written in the representation
given by the product of the ket-vectors $\left|j \right>$ and
$\left|n p_y \right>$, describing the 2D subbands and the Landau
eigenstates, respectively. Here $j=1,2$ numbers the electron subband
of the quantum well, $n$ is the Landau level number, and $p_y$ is
the continuous momentum (the Landau gauge is used). The double
angular brackets in Eq. (2) denote random potential averaging. The
conductivities $\sigma_{\pm}(\omega)$ are expressed in terms of the
functions (2) as
\begin{eqnarray}
\sigma_{\pm}(\omega)= i\frac{e^2n_s}{m \omega} + \frac{e^2}{2 \pi m
\omega} \int d \varepsilon \left[
(f_{\varepsilon}-f_{\varepsilon+\hbar \omega})
Q_{\varepsilon,\omega}^{AR(\pm)} \right. \nonumber \\
\left. +f_{\varepsilon+\hbar \omega}
Q_{\varepsilon,\omega}^{AA(\pm)}-f_{\varepsilon}
Q_{\varepsilon,\omega}^{RR(\pm)} \right].
%3
\end{eqnarray}
In the limit $\omega \rightarrow 0$, Eq. (3) gives the dc
conductivity components $\sigma_{d}={\rm Re}(\sigma_+ +\sigma_-)/2$
and $\sigma_{\bot}={\rm Im}(\sigma_+ -\sigma_-)/2$.

Analytical evaluation of the correlation functions in Eq. (2) can be
done in the case of relatively weak magnetic fields, by using the
self-consistent Born approximation and by expanding the Green's
functions in powers of small Dingle factors. In application to
balanced DQWs, where the subband energy separation is typically
small compared to the Fermi energy, we assume that the difference in
subband populations is small compared to $n_s$. It is justified for our
samples where the Fermi energy (17 meV) is much larger than $\Delta_{12}/2$~=~1.8~meV.
The difference between quantum lifetimes of electrons in the two subbands, $\tau_1$
and $\tau_2$, (and between the corresponding transport times) is
also small, so we neglect it everywhere except the Dingle exponents.
In the first order in the Dingle factors $d_j=\exp (-\pi/\omega_c
\tau_j)$, the generation term acquires the following form:
\begin{eqnarray}
G_{\varepsilon}={\cal G}_{\omega} [ r^{(0)}_{\omega}+
(r^{(0)}_{\omega}-r^{(1)}_{\omega}) g_{\varepsilon} -
r^{(1)}_{\omega} g_{\varepsilon+\hbar \omega}]
\nonumber \\
\times (f_{\varepsilon+\hbar \omega}-f_{\varepsilon}) + \{\omega
\rightarrow -\omega \},
%4
\end{eqnarray}
where
\begin{eqnarray}
{\cal G}_{\omega}=\frac{\pi e^2 E^2 n_s \tau_{tr}}{8 m^2
\omega^2},~~g_{\varepsilon}=\sum_{j=1,2} d_j \cos \frac{2 \pi
(\varepsilon -\varepsilon_j)}{\hbar \omega_c}, \\
r^{(0)}_{\omega}=\sum_{\pm}
\frac{|\lambda^{(\pm)}_{\omega}|^{2}}{1+s_{\pm}},~~
r^{(1)}_{\omega}=\sum_{\pm}\frac{|\lambda^{(\pm)}_{\omega}|^{2}s_{\pm}}{
(1+s_{\pm})^2},
%5,6
\end{eqnarray}
$s_{\pm}=(\omega \pm \omega_c)^2 \tau^2_{tr}$, and $\tau_{tr}$ is
the transport time, common for both subbands in our approximation.

The kinetic equation is solved by representing the distribution
function as a sum $f^{0}_{\varepsilon}+\delta f_{\varepsilon}$,
where the first term slowly varies on the scale of cyclotron energy,
while the second one rapidly oscillates with energy.$^{8}$ The first
term satisfies the equation ${\cal G}_{\omega} r^{(0)}_{\omega}
(f^0_{\varepsilon+\hbar \omega} + f^0_{\varepsilon-\hbar \omega} - 2
f^0_{\varepsilon})=-J_{\varepsilon}(f^0)$. Assuming that the
electron-electron scattering controls the electron distribution, one
can approximate $f^0_{\varepsilon}$ by a heated Fermi distribution,
$f^0_{\varepsilon}=\{1+\exp[(\varepsilon-\varepsilon_F)/T_e]\}^{-1}$.
The electron temperature $T_e$ is found from the balance equation
$P_{\omega}=P_{ph}$, where $P_{\omega}= \int d \varepsilon ~\!
\varepsilon D_{\varepsilon} G_{\varepsilon}(f^0)= {\rm Re} [
|\lambda^{(+)}_{\omega}|^2  \sigma_+(\omega) +
|\lambda^{(-)}_{\omega}|^2 \sigma_-(\omega) ] E^2/4$ is the power
absorbed by the electron system and $P_{ph}=-\int d \varepsilon ~\!
\varepsilon D_{\varepsilon} J_{\varepsilon}(f^0)$ is the power lost
to phonons. The second term satisfies the equation
\begin{eqnarray}
r^{(0)}_{\omega}[\delta f_{\varepsilon+\hbar \omega} +\delta
f_{\varepsilon-\hbar \omega} - 2 \delta f_{\varepsilon}]
-\frac{\delta f_{\varepsilon}}{\tau_{in} {\cal G}_{\omega} }
\nonumber \\
= r^{(1)}_{\omega} [g_{\varepsilon+\hbar
\omega}(f^{0}_{\varepsilon+\hbar \omega}
-f^{0}_{\varepsilon})+g_{\varepsilon - \hbar
\omega}(f^{0}_{\varepsilon-\hbar \omega} -f^{0}_{\varepsilon})],
%7
\end{eqnarray}
where we have used the relaxation time approximation for the
collision integral, $J_{\varepsilon}(\delta f) \simeq -\delta
f_{\varepsilon}/\tau_{in}$, with inelastic relaxation time
$\tau_{in}$. Solution of this equation is
\begin{eqnarray}
\delta f_{\varepsilon} \simeq \frac{\hbar \omega_c}{2 \pi}
\frac{\partial f^{0}_{\varepsilon}}{\partial \varepsilon}
\frac{A_{\omega}}{2} \sum_{j=1,2} d_j \sin \frac{2 \pi (\varepsilon
-\varepsilon_j)}{\hbar \omega_c}, \\
A_{\omega}=\frac{ {\cal P}_{\omega} (2 \pi \omega/\omega_c) \sin (2
\pi \omega/\omega_c)}{1+{\cal P}_{\omega} \sin^2(\pi
\omega/\omega_c) } \frac{r^{(1)}_{\omega}}{ r^{(0)}_{\omega}},
%8,9
\end{eqnarray}
where ${\cal P}_{\omega}= 4 {\cal G}_{\omega} r^{(0)}_{\omega}
\tau_{in}$.

Using the energy distribution function found above, one can
calculate the dc resistivity
$\rho_{xx}=\sigma_{d}/(\sigma_{d}^2+\sigma_{\bot}^2)$:
\begin{eqnarray}
\frac{\rho_d}{\rho_0} = 1 - 2 {\cal T}_{e} g_{\varepsilon_F} +
\frac{\tau_{tr}}{\tau^{tr}_{11}} (d_1^2 + d_2^2)
+ 2 \frac{\tau_{tr}}{\tau^{tr}_{12}} d_1 d_2 \nonumber \\
\times \cos \frac{2 \pi \Delta_{12}}{\hbar \omega_c}  -
\frac{A_{\omega}}{2} \left(d_1^2 + d_2^2+ 2 d_1 d_2\cos \frac{2 \pi
\Delta_{12}}{\hbar \omega_c} \right),
%10
\end{eqnarray}
where $\rho_0=m/e^2 n_s \tau_{tr}$ is the zero-field Drude
resistivity and $\Delta_{12}=\varepsilon_2-\varepsilon_1$ is the
subband separation. The second term, proportional to
$g_{\varepsilon_F}$, describes the SdHO. The third and the fourth
terms describe positive magnetoresistance and the MIS
oscillations,$^{7}$ respectively; these terms are written under
condition of classically strong magnetic fields, $\omega_c \tau_{tr}
\gg 1$. Here, $\tau^{tr}_{11}=\tau^{tr}_{22}$ and $\tau^{tr}_{12}$
are the intrasubband and intersubband transport scattering times,
which contribute to the total transport time according to
$1/\tau_{tr}=1/\tau^{tr}_{11} +1/\tau^{tr}_{12}$. The term
proportional to the oscillating factor $A_{\omega}$ describes
modification of the oscillatory resistivity under photoexcitation.
Another effect of the excitation is the electron heating, which
increases thermal suppression of the SdHO, described by the factor
${\cal T}_{e}=(2 \pi^2 T_e/ \hbar \omega_c)/\sinh(2 \pi^2 T_e/\hbar
\omega_c)$.

The most essential feature of the resistivity given by Eq. (10) is
the presence of the product of the oscillating factors $\cos (2 \pi
\Delta_{12}/\hbar \omega_c)$ and $\sin (2 \pi \omega/ \omega_c)$,
which corresponds to an interference of the MIS oscillations with
the MIRO. The interference takes place because the photoexcitation
involves both subbands [see Eqs. (8) and (9)], so the
scattering-induced coupling between the subbands, which oscillates
as $\cos (2 \pi \Delta_{12}/\hbar \omega_c)$, modifies the amplitude
of the oscillating photocurrent. Usually, the microwave quantum
energy is smaller than the subband separation, and the
magnetoresistance shows fast MIS oscillations modulated by a slow
MIRO component $\propto -\sin (2 \pi \omega/ \omega_c)$.

In our calculations, we have used Eq. (10) with $d_1=d_2=\exp
(-\pi/\omega_c \tau)$ and $\tau^{tr}_{12}=\tau^{tr}_{11}
=2\tau_{tr}$, which is a good approximation for balanced DQWs.$^7$
We have taken into account heating of electrons by the field and
dependence of the characteristic times $\tau$, $\tau_{tr}$, and
$\tau_{in}$ on the electron temperature $T_e$. The latter was
determined by using the collision integral for interaction of
electrons with both deformation and piezoelectric potentials of
acoustic phonons. The inelastic scattering time $\tau_{in}$ is
assumed to scale with the temperature as $T_e^{-2}$, with
$\hbar/\tau_{in}=4$ mK at $T_e=1$ K, according to the theoretical
estimates for electron-electron scattering$^{8}$ applied to our
samples. The temperature dependence of the transport time
$\tau_{tr}$ and quantum lifetime $\tau$ of electrons in our samples
has been empirically determined by analyzing temperature dependence
of zero-field resistance and MIS oscillation amplitudes.$^{7}$ The
microwave electric field $E \simeq 2.5$ V/cm is estimated by fitting
calculated amplitudes of magnetoresistance oscillations to
experimental data; the corresponding electron heating is in
agreement with the observed suppression of the SdHO amplitudes.

The results of the calculations, shown in the lower parts of Figs. 1
- 3, are in good agreement with experiment, and capture all
qualitative features of the resistance dependence on the magnetic
field, temperature, radiation power, and frequency. Note that the
heating of electrons by microwaves is not strong at $\omega_c \simeq
\omega$ (see inset to Fig. 2) because of radiative broadening of the
cyclotron resonance$^{10,11}$ in our samples. The only feature that
is not described by the theory is the observed reduction in the MIS
oscillation frequency$^7$ at $B \gtrsim 0.25$ T, which is possibly
related to a decrease in the subband separation owing to enhanced
Coulomb correlations and/or modification of screening in magnetic
fields.

Understanding the MIRO physics requires comparison of measured
photoresistance with the results provided by existing theories.
Among the physical mechanisms$^{12}$ responsible for the MIRO, the
inelastic mechanism,$^{8,12}$ describing the oscillations as a
result of microwave-induced change of the isotropic part of electron
distribution function, is predicted to dominate at moderate
radiation power, owing to a large ratio of $\tau_{in}/\tau$, which
is of the order $10^2$ under typical experimental conditions,
including our experiment. The observed insensitivity of the MIRO to
the polarization of incident radiation, and $T^{-2}$ scaling of the
oscillatory photoresistance amplitude$^{11}$ are in favor of the
inelastic mechanism. We have shown that application of the basic
principles of the inelastic mechanism theory$^8$ to the systems with
two-subband occupation
%allows a broad comparison of theory and
%experiment for various temperatures, frequencies, and intensities of
%the radiation,
confirms the reliability of theoretical estimates for the inelastic
relaxation time and leads to a satisfactory explanation of the new
features reported in this paper.

In conclusion, we have presented experimental and theoretical
studies of oscillatory magnetoresistance in DQWs under microwave
irradiation. The new kind of interference oscillations observed in
these systems appears because the photoinduced part of the electron
distribution, which oscillates as a function of microwave frequency,
is modified owing to subband coupling and becomes also an
oscillating function of the subband separation.

This work was supported by CNPq, FAPESP (Brazilian agency), 
COFECUB-USP (Uc Ph 109/08), and with microwave facilities from ANR
MICONANO.

\end{document}